\documentclass[a4paper,fleqn]{cas-sc}

\usepackage[authoryear,longnamesfirst]{natbib}
\usepackage{subfig}
\usepackage{multirow}
\usepackage{todonotes}

\def\tsc#1{\csdef{#1}{\textsc{\lowercase{#1}}\xspace}}
\tsc{WGM}
\tsc{QE}

\newcommand{\finding}[3]{\noindent\\\fbox{%
\parbox{\dimexpr%
    \linewidth-2\fboxsep-2\fboxrule}%
    {\textbf{Finding #1: \underline{#2}} #3}}\\}

\begin{document}
\let\WriteBookmarks\relax
\def\floatpagepagefraction{1}
\def\textpagefraction{.001}

\shorttitle{}    

\shortauthors{Daniel et al.}  

\title [mode = title]{Reference Architecture for Metadata-driven Services to Promote Reusability in Software Systems}  



%

\author[1]{João F. L. Daniel}[orcid=0000-0002-2877-9509]
\cormark[1]
\ead{joao.daniel@student.unibz.it}
\credit{}

\author[2]{Bruno P. Romano}[orcid=0009-0000-4305-1295]
\ead{brunoromano@ime.usp.br}
\credit{}

\author[3]{Xiaofeng Wang}[orcid=0000-0001-8424-419X]
\ead{}
\credit{}

\author[1]{Andrea Janes}[orcid=0000-0002-1423-6773]
\ead{andrea.janes@unibz.it}
\credit{}

\author[1]{Eduardo M. Guerra}[orcid=0000-0001-5555-3487]
\ead{eduardo.guerra@unibz.it}
\credit{}

\affiliation[1]{organization={Free University of Bozen-Bolzano},
            addressline={Piazza Università, 1}, 
            city={Bozen-Bolzano},
            postcode={39100}, 
            state={Trentino Alto-Adige},
            country={Italy}}

\affiliation[2]{organization={University of São Paulo},
            addressline={Rua da Reitoria, 34}, 
            city={Cidade Universitária},
            postcode={05508-220}, 
            state={São Paulo},
            country={Brazil}}

\affiliation[3]{organization={Lappeenranta–Lahti University of Technology},
            addressline={Yliopistonkatu 34}, 
            city={Lappeenranta},
            postcode={53850}, 
            state={Lappeenranta},
            country={Finland}}

\cortext[1]{Corresponding author}



\begin{abstract}
    Service-based Architectures place reusability among their central design goals, yet structural heterogeneity across clients often drives the creation of services with similar functionalities, undermining system evolution and maintainability. In this work, we address this issue by focusing on validated architectural artifacts that bound to a limit the number of replicated services. We do so by proposing and validating a reference architecture that employs metadata as the core mechanism to promote service reusability, embracing heterogeneous data. The proposed RA is designed based on a pattern language with the same purpose, and it is evaluated by combining two well-established methods for RA evaluation: scenario-based evaluation and case studies with real-world systems. The triangulation of these methods' results demonstrated that, during the system's evolution, the most common change types the RA incurs are either no change or less impactful ones, like configuration changes or the addition of a pluggable class.
\end{abstract}

\begin{highlights}
\item A reference architecture for metadata-driven service reusability is proposed
\item The reference architecture is grounded in a peer-reviewed pattern language
\item Metadata enables a single service to absorb structurally heterogeneous clients
\item Architecture evolution favors configuration and pluggable changes over code changes
\item Four real-world systems empirically corroborate the reference architecture's design
\end{highlights}
\begin{keywords}
 service architectures \sep architecture model \sep software architecture patterns \sep API design \sep reusable services \sep reference architecture \sep metadata
\end{keywords}

\maketitle

\section{Introduction}

Service-based Architectures (SbA) has established itself as the standard architecture paradigm for designing software systems. It encompasses the architectural styles of Service-Oriented Architecture (SOA) and Microservices Architecture (MSA) because of the shared principle: exposing the system's capabilities through well-defined API contracts, favoring decoupling and reuse \citep{sommerville2011software, papazoglou2008web, gorton2006essential}. Reusability is among these styles' motivations to improve development productivity and cohesion in the architecture.

Nonetheless, despite reuse being a goal of SbA, achieving it in practice has challenges. Desirable qualities like modularization and fine-grained boundaries also create conditions that tend to replicate access to similar functionality due to differences in data structure or semantics \citep{carvalho2019extraction}. That is because the straightforward solution is to create dedicated services for each case of the heterogeneous ecosystem of clients. But, over time, that incurs costly maintenance; after all, the same capability is spread across multiple similar implementations.

Addressing this issue requires architectural mechanisms that allow a single service to embrace structural heterogeneity without compromising coherence or extensibility. One approach to that is to design the API contracts to allow any data structure, as long as it comes with a layer of describing metadata. This idea can be seen in API patterns such as \textsc{Wish List} \citep{Zimmermann2023PatternsExchanges} and in a dedicated pattern language for metadata-driven service reusability \citep{LinoDaniel2024PatternAPIs, daniel2025incrementing}. Even though a pattern language communicates solutions to a variety of problems, it might not fit the needs of practitioners seeking solutions for a broader scope, in which multiple such problems occur simultaneously.

Reference architectures (RAs) address this gap: by assembling components that play the roles described in the patterns, a pattern-driven RA offers a more concrete and actionable guide, standing out especially when it comes to mapping the patterns into components and the interactions between these components. Nonetheless, RAs are defined in a level of abstraction that still enables their application to multiple different concrete scenarios \citep{Guerra2015RelatingArchitectures}. Software RAs are used in a wide range of domains, and their evaluation through scenario-based and case studies is a well-establish practice \citep{kazman2002scenario, Moreira2026}.

The goal of this work is to promote service reusability in the context of service-based architectures via an architectural design approach, aiming at providing a structure that bounds the amount of services with replicated or similar functionalities. From this goal, we distilled the following Research Questions (RQ):

\begin{description}
    \item[RQ1] In the context of SbA, how can a RA bound the amount of services with replicated or similar functionalities, yielding service reusability?\\
    \textsc{rationale} $\Rightarrow$ \textit{proposing a RA for service reusability is an approach that connects components in collaboration towards well-defined requirements, at the same time that is abstract and flexible enough to be used in multiple concrete scenarios};

    \item[RQ2] To what extent does this RA bound the amount of services with replicated or similar functionalities?\\
    \textsc{rationale} $\Rightarrow$ \textit{although the RA is design to fulfill its requirements, it is important to assess more specifically how well each feature is addressed by the proposed reference architecture, both conceptually and in realistic conditions};
\end{description}

\section{Reusability in Service-based Architectures}

SbA encompasses various architectural styles, more distinctly \textit{Service-Oriented Architecture (SOA)} and \textit{Microservices Architecture (MSA)} -- although there are other well-known approaches, like \textit{Self-Contained Systems} and \textit{Serverless} or \textit{Function-as-a-Service}. Regardless, the central aspect of these styles is the concept of \textbf{service}, which, among multiple definitions, is an approach to expose a system's capability through a well-established API contract in a reusable and loosely coupled way \citep{sommerville2011software, papazoglou2008web, gorton2006essential}.

Despite SbA having reuse as a key concern, there are different interpretations of its meaning \citep{oliveira2018soa}; some of them are more centered in structural and infrastructural aspects, and others perceive service reuse as the ability to be embedded in different business processes or as a means to achieve higher software quality, regarding improving cohesion and reducing replication. The reasons to seek service reuse also vary, ranging from engineering-related improvements like development agility and productivity to architectural qualities like decoupling and flexibility \citep{oliveira2018soa}. In the same work, they also list strategies to achieve service reuse, which highlights the impact of domain modeling around services and following service-oriented standards \citep{oliveira2018soa}.

The adoption of MSA, especially because of the migration to an architecture centered in service, brings to light the need for flexible services to enable reuse \citep{carvalho2019extraction}. This process enables the redesign of APIs, and one approach to favor reuse is to slightly broaden their capabilities or shared data, making it suitable for a wider range of clients \citep{carvalho2019extraction}. Although this is a liability, there are documented, well-known practices to deal with it, like the API patterns \textsc{Wish List} and \textsc{Wish Template} \citep{Zimmermann2023PatternsExchanges}, that use metadata to enable clients to select specific fields on the fetched data.

Lastly, service reusability is not a synonym of self-adaptability: the former is a design-time concern that allows a service to be part of different business contexts; the latter refers to the runtime capability of responding to changes in the context and in response to execution events \citep{mendoncca2018generality}, like a surge in demand or an error.
\section{Research Method}\label{sec:ra-design}

We designed our research method based on the two research questions presented in this study, dividing it into two phases. In the first phase, addressing RQ1, we designed an RA whose goal is to promote reusability in SbA, described in Subsec. \ref{subsec:refarch-design}. In the second phase, focusing on RQ2, the designed RA was evaluated based on architectural scenarios and case studies, as described in Subsec. \ref{subsec:eval-design}.


\subsection{Reference Architecture Design}\label{subsec:refarch-design}

We distilled \textbf{RQ1} and defined that our proposed RA should adhere to the following functional requirements and quality attributes:

\begin{itemize}
    \item Functional Requirements (FR):
    \begin{description}
        \item[(FR-1) Heterogeneous Data Ingestion:] rather than multiple dedicated operations per client type or domain object, the service must expose in its API contract a single ingestion operation capable of handling heterogeneous data;
        
        \item[(FR-2) Contextual Data Characterization:] the service must expect an additional information layer (metadata) to characterize the heterogeneous data, in structure, semantics, or both;
        
        \item[(FR-3) Data-driven Processing Adaptation:] in design time, there is not enough information on how to process ingested data; hence, the service must adapt the processing steps to the data and metadata received;

        \item[(FR-4) Provider-side Characterization Bridging:] as the API contract's metadata requirement might not be fulfillable in many cases, the RA must provide the means to satisfy on the clients' behalf;
        
        \item[(FR-5) Structural Mediation Between Incompatible Parties:] clients might be interested in the capabilities of a legacy (sub)system having either parts unable to adapt to each other; the architecture must support the mediation between these parts based on metadata describing the transformations between expected and provided structures;
        
        \item[(FR-6) Dynamic External Service Integration:] the architecture must support a seamless approach to integrate $3^{rd}$-party service providers, via self-registration, describing in metadata the communication to be established;
        
        \item[(FR-7) Result Aggregation and Flexible Retrieval:] in the case of data being stored and processed to be served later, the architecture must provide solutions to that. The service must expose a flexible reading operation, in which the data is responded to with the accompanying metadata so that the client can interpret it, and a normalized reading operation, in which the metadata is used by the service to transform the data into the defined shape.
    \end{description}

    \item Quality Attributes (QA):
    \begin{description}
        \item[(QA-1) Service Reusability Across Structural Heterogeneity:] the service itself must have the conditions to homogeneously respond to the heterogenous demands of its clients, without compromising its functional capabilities;
        
        \item[(QA-2) Seamless Extensibility of Providers:] a provider is understood as either a client that provides data through the API contract, or another service that provides its capabilities with its own API contract; the service must support accommodating a new provider requiring the least significant design impact possible, ideally having no impact;
        
        \item[(QA-3) Seamless Extensibility of Consumers:] the service must support accommodating a new consumer requiring the least significant design impact possible, ideally having no impact;
        
        \item[(QA-4) Seamless Extensibility of Processors:] the architecture must support accommodating a new processor, requiring the least significant design impact possible, ideally having no impact.
    \end{description}
\end{itemize}

The RA was designed following the pattern-based method for creating reference architectures \citep{Guerra2015RelatingArchitectures}. Patterns excel in communicating a contextualized solution to a problem, and pattern languages do the same to various related problems within the same context. Nonetheless, there is no single way to implement a pattern, much less a combination of different patterns. Meanwhile, reference architectures assemble software components that play the roles proposed by the patterns, providing a clearer view of how to combine the solutions.

We selected a pattern language created in previous work as a starting point: it is intended for metadata-driven services to promote reusability \citep{LinoDaniel2024PatternAPIs}, which we summarized in Subsec. \ref{subsec:plang}. We mapped its patterns into components of an architecture based on the role the pattern's solution played. In some cases, a single pattern led to more than one component -- such as the \textsc{Plug the Processors In} pattern, from which we derived the \textsc{Processing Orchestrator} and the \textsc{Processor} interface; in other cases, a single component embedded solutions from different patterns -- as it happened with the \textsc{Processor Proxy} component that brings together the solutions of both patterns, \textsc{Plug the Processors In} and \textsc{Proxy the Plugin}.

Once the patterns had been mapped into a comprehensive architecture, we identified three distinct concerns. So, even though they are related and can provide synergy within an architecture, they could be presented separately. Based on that, we separated the proposed RA into the following three parts so as to reduce the cognitive load required to understand each one:

\begin{enumerate}
  \item \textbf{Data Ingestion and Processing}: here lies the core of the pattern language, where metadata is used to guide the processing of data sent from external providers;
  
  \item \textbf{Moderation}: the common factor of this concern is a structural feature entailed by the roles played by the patterns: to serve as supporting intermediates in the communication between systems, typically bridging internal ones to external ones;
  
  \item \textbf{External Consumption}: it can be seen as a specialization of some solutions in the "moderation" slice when enhancing functionalities with $3^{rd}$-party services.
\end{enumerate}

The proposed RA comprises these concerns, but they are not mutually exclusive; rather, they are composable. One challenge entailing the \textsc{Data Ingestion and Processing} is the compliance of the clients, and that motivates part of the \textsc{Moderation} concern. Similarly, when processing is required for the \textsc{External Consumption} or \textsc{Moderation} concerns, structures of the \textsc{Data Ingestion and Processing} can be employed.

\begin{figure}
    \centering
    \includegraphics[width=0.65\linewidth]{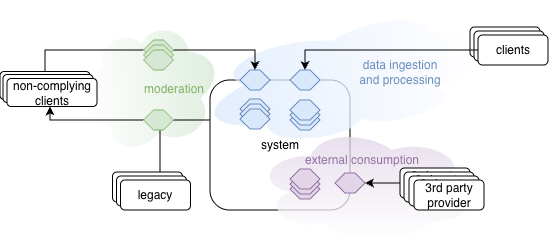}
    \caption{An overview of the proposed RA with highlights of its three concerns}
    \label{fig:concerns-overview}
\end{figure}

Fig. \ref{fig:concerns-overview} illustrates an overview of the proposed RA highlighting these concerns.

\subsection{Evaluation Design}\label{subsec:eval-design}

\citet{Moreira2026} shows that evaluating a RA is a complex task, often requiring a combination of evaluation methods. Our evaluation was driven by some overarching features, such as suitability for the requirements and applicability to real software systems. We designed our evaluation methodology by mixing two well-established evaluation methods for reference architectures: scenario-based assessment and case study.

On the one hand, scenario-based evaluation consists of mapping the requirements and quality attributes into scenarios to illustrate how the objects -- in this case, the elements of the RA -- perform in a meaningful and representative context \citep{kazman2002scenario}. On the other hand, a case study is an empirical method for contextual investigation, characterized by the realism of its findings and by its flexibility in design \citep{Yin2009CaseMethods,Runeson2009GuidelinesEngineering,Wohlin2022IsGuidance}. Such flexibility allows it to be combined with scenario-based assessment, resulting in a stronger grounding of the evaluation in existing applications and reality.

Our concrete evaluation design mixed these methods and assessed the RA with the triangulation of their results. We started the scenario-based method by defining the scenarios based on the functional requirements and quality attributes presented in Subsec. \ref{subsec:refarch-design}. There are three scenarios with realistic elements, their interaction with each other, and a set of change-provoking stimuli. Then, we defined a taxonomy of changes with an increasing impact required to resolve the provoking stimuli. Next, we conducted a Change Impact Analysis \citep{Zhao2002ChangeEvolution} of the RA, where we applied the RA's components as the reaction to the provoking stimuli of each scenario and analyzed it according to the taxonomy of change.

\begin{figure}
    \centering
    \includegraphics[width=\linewidth]{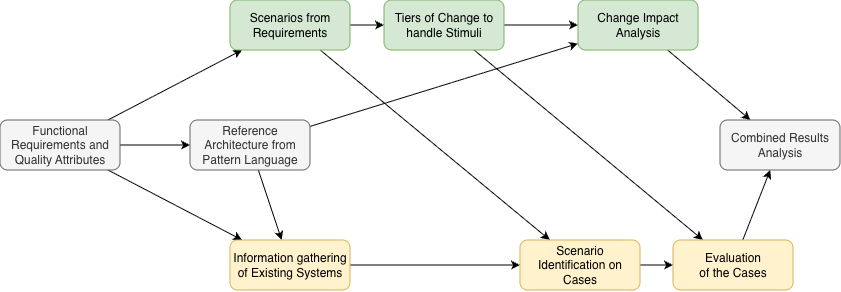}
    \caption{Concrete steps taken in the designed methods for this work (in green, scenario-based method; in yellow, case study method)}
    \label{fig:methods-steps}
\end{figure}

In parallel, using the case study method, we gathered information on existing systems to use as a case study, following purposive sampling \citep{Baltes2022SamplingGuidelines}. We selected four systems based on the following criteria: having similar requirements, implementing our proposed RA partially or fully, and having maintainers available for clarifications. Then, we observed the scenarios in each case and evaluated the role the RA elements in the implementation played in resolving the scenarios' provoking stimuli. Finally, for each scenario, we compiled the results from each method. Fig. \ref{fig:methods-steps} illustrates these steps, which we highlighted in green, the scenario-based method, and in yellow, the case study method.

\subsection{Scenario-based method}\label{subsec:scenarios}

Scenario-based is one of the most adopted methods for evaluating references architectures \citep{Moreira2026}. It is used in other works as a means to reveal the reference architecture's behavior under specific circumstances, regardless of the domain \citep{liu2018towards, bocciarelli2023tosca, dirin2023security}.

Given the reference architecture's focus on the evolution of microservice architectures, the scenarios target common actions in this architectural style, such as adding service consumers and providers. We created three scenarios in which the proposed RA should excel. 

Each scenario provides a contextual description and contains stimuli that the architecture must handle. Furthermore, a scenario is described with a mapping into more than one of the RA's functional requirements (FRs) or quality attributes (QAs) it addresses and illustrated by a components diagram.

\begin{description}
    \item[Scenario 1] \textit{Metadata-based Platform onboarding New Providers.} A platform is built to receive and process data from a variety of independent providers. These providers operate in different domains, were built at different times, and have no coordination with each other; each sends data in its own structure and follows its own conventions. Some providers were able and willing to describe their data when sending it; others lacked this capability or would not change. The platform serves all of them through a unified interface, processing each provider's data correctly and independently of others. The stimuli:
    \begin{itemize}
        \item a new provider that describes its own data is onboarded to the platform;
        \item a legacy provider that cannot describe its data is onboarded to the platform;
        \item both send data concurrently.
    \end{itemize}
    \begin{description}
        \item[FRs] FR-1, FR-2, FR-3, FR-4
        \item[QAs] QA-1, QA-2, QA-4
    \end{description}

    \begin{figure}
        \centering
        \includegraphics[width=0.35\linewidth]{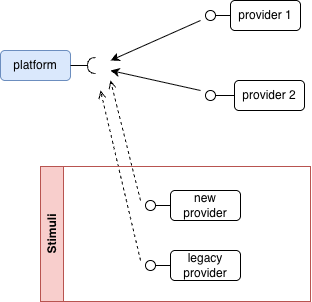}
        \caption{Scenario 1: Metadata-based Platform onboarding New Providers}
        \label{fig:placeholder}
    \end{figure}

    \item[Scenario 2] \textit{Service Connecting Hetereogeneous Consumers and Providers.} A service has been in stable operation with an established set of providers and consumers. The system works, contracts are settled, and no party is willing to change their interface. Then the environment shifts: a new provider emerges with a data structure unlike any currently supported; separately, a new consumer wants to interact with an existing provider but speaks a structurally different language. Neither the new consumer nor the existing provider can adapt to the other. The system must absorb both without disturbing what already works. The stimuli:
    \begin{itemize}
        \item the new provider is introduced to the system;
        \item the new consumer is introduced, targeting an existing provider;
        \item the system continues operating.
    \end{itemize}
    \begin{description}
        \item[FRs] FR-3, FR-5, FR-6, FR-7
        \item[QAs] QA-1, QA-2, QA-3
    \end{description}

    \begin{figure}
        \centering
        \includegraphics[width=0.35\linewidth]{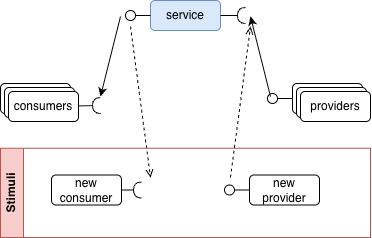}
        \caption{Scenario 2: Service Connecting Heterogeneous Consumers and Providers}
        \label{fig:placeholder}
    \end{figure}

    \item[Scenario 3] \textit{Enriching Capabilities with $3^{rd}$-party.} A service needs to enrich its core functionality by delegating certain operations to external $3^{rd}$-party providers. Multiple providers offer the same service category, but each has its own invocation protocol and returns results in its own format. The system must be able to work with any of them interchangeably and accommodate new ones as they emerge, without components that request enrichment needing to know which provider they are talking to. The stimuli:
    \begin{itemize}
        \item two external providers offering the same category of functionality are registered with the system;
        \item a consuming component requests the functionality without specifying a provider;
        \item a new external provider is introduced; the consuming component remains unchanged.
    \end{itemize}
    \begin{description}
        \item[FRs] FR-2, FR-5, FR-6, FR-7
        \item[QAs] QA-1, QA-2, QA-4
    \end{description}

    \begin{figure}
        \centering
        \includegraphics[width=0.35\linewidth]{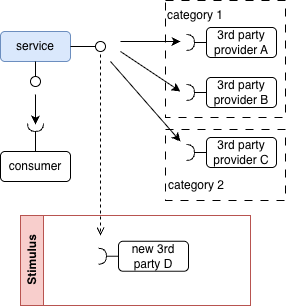}
        \caption{Scenario 3: Enriching Capabilities with $3^{rd}$-party}
        \label{fig:placeholder}
    \end{figure}
\end{description}

With these scenarios, we manage to cover all the defined functional requirements and quality attributes for the RA. Table \ref{tab:reqs-per-scenario} assists in visualizing that, describing the FRs and QAs of each scenario.

\begin{table}
    \centering
    \begin{tabular}{l|ccccccc|cccc}
                    & FR-1 & FR-2 & FR-3 & FR-4 & FR-5 & FR-6 & FR-7 & QA-1 & QA-2 & QA-3 & QA-4\\\hline
        Scenario 1  &  X   &   X  &   X  &  X   &      &      &      &   X  &   X  &      &   X \\\hline
        Scenario 2  &      &      &   X  &      &   X  &  X   &   X  &   X  &   X  &   X  &     \\\hline
        Scenario 3  &      &   X  &      &      &   X  &  X   &   X  &   X  &   X  &      &   X
    \end{tabular}
    \caption{Requirements (FR and QA) covered by Scenario}
    \label{tab:reqs-per-scenario}
\end{table}

In each scenario, we used architecture change impact analysis \citep{Zhao2002ChangeEvolution} to evaluate the RA's capability by considering the type of change required for each component of the architecture. Based on the mechanisms "slicing" and "chopping" \citep{Zhao2002ChangeEvolution}, we proposed a taxonomy of changes that applied to the abstractions in our RA, as follows:

\begin{itemize}
    \item \textbf{No change (NoC)}: No change is needed in a component;
    \item \textbf{Configuration change (CfC)}: A change needs to be performed in the configuration;
    \item \textbf{(Pluggable) Class addition (pCA)}: A new class needs to be added to the system, but without changing the code of existing ones;
    \item \textbf{External Subsystem addition (ESA)}: A new subsystem needs to be added to the system, but in an external fashion, not requiring changes to the existing subsystems;
    \item \textbf{Source code change (CSC)}: The code of the component needs to be changed, requiring it to be repacked and redeployed;
    \item \textbf{API Contract change (ACC)}: Change in API, requiring changes to both sides of the communication -- the provider and the consumers.
\end{itemize}

The goal of the reference architecture is to avoid higher-impact changes in these scenarios and favor more lightweight ones, such as metadata configuration changes. Adding components is also acceptable, as it still preserves the open-closed principle. The need for source code and API changes would reveal that the reference architecture still does not support smooth architectural evolution, since these changes have a higher impact and are more expensive to implement.

\subsection{Case study method}


\begin{description}
    \item[Open Data Hub] is an open-source platform that was created to promote innovation in South Tyrol, Italy. It does so by being an open data platform that exposes its capabilities through an API, fed by numerous partners in the region. The API acts as an abstraction layer for its microservice-based architecture, centered on two main components: the Ingestion Service and the Aggregated Database. Orbiting the former, there are numerous $3^{rd}$-party data providers, some autonomously sending their data, others requiring dedicated data fetchers. When data is sent to Ingestion, it might be accompanied by metadata descriptors and locally stored. According to their processing policy, these raw data records are asynchronously sent to the Transforming Service that orbits the Aggregated Database. Its responsibility is to handle the heterogeneity in data shapes from the providers and maintain a consistent, normalized shape in the Aggregated Database. These transformations are made based on the metadata when present.

    \item[Catch\&Solve] is a software-based startup in South Tyrol, Italy, that provides quality-checking services to other software-based companies \citep{Silva2024UsingStart-up}. The solution consists of a range of Check Agents, each for a different check type -- for instance, a toolkit that is built with Android apps and sends checks of exceptions happening at runtime.  These checks are assigned a check-ID and sent to the Ingestion API, a microservice that receives and persists the data. Eventually, the Dashboard Backend reads the raw data, processes it according to the service mapped to the check-ID metadata, and then aggregates it into reports. A user can access such reports in a web-based dashboard.

    \item[Digi Dojo] is a platform implemented to support a research project aiming at creating virtual workspaces for early-stage startups. Its implementation has four domain microservices: "Startups and Users", "Virtual Spaces", "Tasks and Calendars", and "Assistant", which is dedicated to research support. Additionally, there is the "Gateway Router and Balancer" that has an infrastructure-related role of abstracting the internal separation of microservices and a meta-domain responsibility to feed the "Assistant" with the logs of the incoming requests. The Assistant supports research queries based on the logs sent and uses three tiers of pluggable processors.

    \item[Metrics Platform] is an open-source platform that supports the collection of metrics for architectural analysis of software systems. It was developed as a contribution of research for a master's degree. Once a system is registered in the platform, one can configure metrics to be analyzed. The platform supports dynamic registration of external metric collectors, extending its capabilities. Its architecture is centered on the "Manager" microservice acting as a hub for the collectors, who are responsible for registering both systems to be analyzed and metric collectors to do the work.

\end{description}

An important aspect is that the cases are related to the scenarios presented in Subsec. \ref{subsec:scenarios}. In each case, we can see some of the scenarios implemented. Table \ref{tab:scen-summary} presents which scenario happens in each case.

\begin{table}
    \centering
    \begin{tabular}{l|ccc}
                         & Scenario 1 & Scenario 2 & Scenario 3\\\hline
        Open Data Hub    &     X      &    X       &           \\\hline
        Catch\&Solve     &     X      &    X       &           \\\hline
        Digi Dojo        &     X      &            &           \\\hline
        Metrics Platform &            &            &     X
    \end{tabular}
    \caption{Scenarios observed in each case}
    \label{tab:scen-summary}
\end{table}

\section{Reference Architecture}

In this section, we present a reference architecture that aims at answering this work's \textbf{RQ1}. We designed a reference architecture based on a pattern language with similar scope and goals.

The proposed RA is organised around the concerns "Data Ingestion and Processing", "Moderation", and "External Consumption", each in a dedicated segment of this subsection. We leveraged UML to represent the proposed RA, but with two simplifications over the use of stereotypes: the \textbf{Flexible Provided Interface} and the \textbf{Flexible Requested Interface}. They are explained in the Appendix \ref{appendix:uml-ext}.

\subsection{Base Pattern Language}\label{subsec:plang}

The pattern language we used was published in two complementary segments in continuation of its definition over two years. In total, there are 12 patterns, where the first segment with 8 patterns was published in \citep{LinoDaniel2024PatternAPIs}, and the second with the remaining 4 in \citep{daniel2025incrementing}.


The summary we present here contains all 12 patterns, each in a shorter version of its text that contains three phrases presenting its core features: the context of occurrence, the problem it solves -- shaped as a question --, and the solution statement.

\begin{itemize}
    \item \textbf{Ingestion, Processing and Returning}
    \begin{description}
        \item[\textsc{Flexibilize the Ingestion}    ] A service is constrained to handle heterogeneous data in its API by its various clients. \textit{How do we define its API contract to ensure it can be used in multiple different contexts?} Define an API Contract that, as parameters to its operation, accepts multiple data structures.
        
        \item[\textsc{Flexibilize the Return}       ] An entity of a service's domain is represented in various structures for each record. \textit{How to centralize access to the same type of entity, whose instances might have different structures?} Provide a single operation that returns the entity in its most suitable format for its context, and add metadata that allows the client to interpret its particular structure. 
        
        \item[\textsc{Plug the Processors In}       ] A service accepting heterogeneous data needs to process it. \textit{How to process data that can vary in structure, domain, or nature, and can evolve in time?} Design the data processing to accept plugins, each partially processing the input.
        
        \item[\textsc{Enrich with Metadata}         ] A service with a flexible ingestion that deals with clients in different contexts. \textit{How can a service understand semantic and structural aspects of the data without implicitly making assumptions?} Adopt metadata as a way to enrich the messages received.
    \end{description}

    \item \textbf{Moderating $3^{rd}$-Party}
    \begin{description}
        \item[\textsc{Proxy the Plugin}             ] In a plugin-based architecture, one of the plugins has specific quality attributes that differ from the rest. \textit{How to enable the integration of a plugin processor, selected based on the request metadata, that requires specific qualities?} Have the processor as an external component, and create the local plugin as a proxy, which will be invoked depending on the metadata associated with the request. 
    
        \item[\textsc{Adapt the Ingestion}          ] Metadata is adopted, but a client is unwilling or unable to send it to the service. \textit{How to provide the metadata about referring to an API Consumer unwilling or impossible to change?} Adapt the interaction between API Provider and API Consumer with a component that is capable of providing the API Consumer’s metadata on its behalf.
        
        \item[\textsc{Consume as Plug-and-Play}     ] A system that relies on 3rd party providers for a service it needs. \textit{How can a client dynamically consume a service offered by multiple providers in different formats?} Enable the dynamic integration of new service providers by using metadata to describe how the service should be invoked and how their output should be interpreted. 
        
        \item[\textsc{Mediate with Metadata}        ] Service provider do not match exactly the structural expectations of their potential consumers. \textit{How can a service and various consumer components be seamlessly integrated when there are unchangeable and incompatible message structures?} Mediate the interaction between service and consumers with a component that uses metadata to guide the transformation of structures. 
    \end{description}

    \item \textbf{Tuning Metadata}
    \begin{description}
        \item[\textsc{Configure Meta during Deploy} ] A metadata layer is added on top of the heterogeneous data, so that the service and its clients can communicate clearly. Also, there are some pieces of metadata that are owned by the service instead. \textit{How to provide metadata about the heterogeneous data in the received message?} Design the API Provider to accept deploy-time configuration for the metadata.
        
        \item[\textsc{Agree on Meta Structure}      ] Metadata was adopted to overcome the challenges of heterogeneous data. \textit{How to avoid repeating in the metadata layer the uncertainties of the heterogeneity in the data layer?} Seek agreement on the structure of the metadata among the different agents: senders and receivers of data.
        
        \item[\textsc{Configure Meta in Runtime}    ] A metadata layer is added on top of the heterogeneous data, so that the service and its clients can communicate clearly. Also, each client is known for having a stable data structure. \textit{How to provide metadata about the heterogeneous data in the received message?} Include a metadata configuration service in the API Contract of the API Provider, so API Consumers configure their metadata once to be used onwards.
        
        \item[\textsc{Embed Metadata}               ] A metadata layer is added on top of the heterogeneous data, so that the service and its clients can communicate clearly. \textit{How to provide metadata about the received heterogeneous data?} Expect metadata as part of the message received.
    \end{description}
\end{itemize}

\begin{figure}
    \centering
    \includegraphics[width=0.5\linewidth]{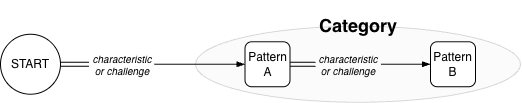}
    \caption{Schematics for the Pattern Language navigation map}
    \label{fig:navmap-schematics}
\end{figure}

\begin{figure}
    \centering
    \includegraphics[width=\linewidth]{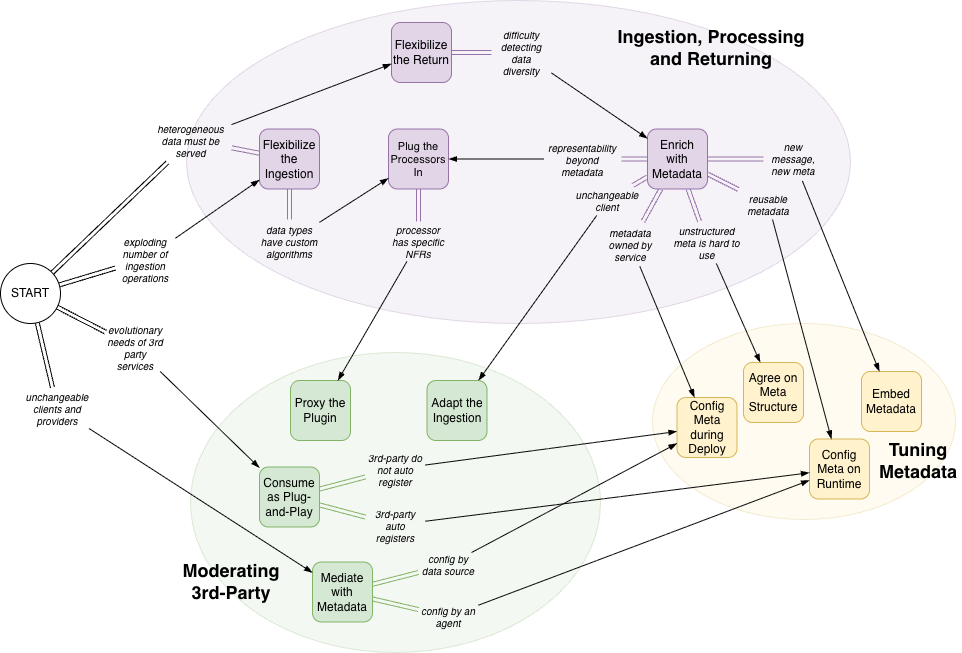}
    \caption{The navigation map for the pattern language used as a base for the creation of the reference architecture}
    \label{fig:navmap}
\end{figure}

The pattern language also contains a navigation map that relates different patterns in a chain of refinement and composition. More specifically, the map displays the patterns and some of their resulting characteristics or challenges, which are addressed by another pattern, leading to a progressive adoption scheme for the pattern language. Fig. \ref{fig:navmap} illustrates this map, according to the schematics of Fig. \ref{fig:navmap-schematics}. Its reading begins in the START circle and proceeds by navigating a characteristic/challenge that leads to a pattern; from there, the navigation follows the same dynamic, selecting the next characteristic/challenge to address and navigating its line. The patterns in the map are also grouped into categories by the type of problem they are solving: "Ingestion, Processing and Returning", "Moderating $3^{rd}$-Party", "Tuning Metadata".

As an example of use, one with "unchangeable clients and providers" issue in their system, starting from the START the map leads to implementing \textsc{Mediate with Metadata}; next, deciding to approach the "configuration by data source", the map leads to implementing \textsc{Config Meta during Deploy}.

\subsection*{Concern 1: Data Ingestion and Processing}

A Data Ingestor microservice interacts with a wide range of Data Providers. The ingestor serves a capability that can input a variety of data structures. Instead of offering dedicated operations for each data structure, the ingestor leverages metadata as an approach to deal with different data homogeneously.

The metadata can come from three sources or periods of time:
\begin{enumerate}
  \item the maintainers of the ingestor can provide metadata during deployment;
  \item the maintainers of a data provider can configure their metadata during the runtime of the ingestor;
  \item the data sent by the data providers can embed its metadata.
\end{enumerate}

To support the (2) approach, the ingestor offers a “configure metadata” operation publicly. The Metadata Configuration Manager handles the call and stores the valid metadata within the Metadata Repository. This repository abstracts the persistence and aggregates the just-received metadata with its existing content, which encompasses the metadata set using (1) flow.

Data ingestion occurs via two operations offered by the Flexible Data Ingestor object within the ingestor microservice: one for ingesting data with embedded metadata, and another for ingesting data without metadata. Regardless of which operation a data provider calls, the flexible ingestor triggers data processing. It calls the Processing Orchestrator, which receives the data and combines it with all referring metadata (the pieces that might have come with the data, and those stored in the Metadata Repository).

Data processing is split into small units, each implementing the Processor interface. Some processors might run locally within the same operating system (OS) process as the Ingestor microservice – these are called Local Processors. In some cases, it might be necessary to implement the processor as a remote OS process – e.g., when it requires a different programming platform or when non-functional requirements apply – or it might already be implemented on a remote system. In these cases, the Processor interface is implemented as a Proxy object that redirects calls to the remote system. They are named External Service Proxy and System Service Proxy, respectively, when the implementation is 3rd-party software and when it is part of the same system.

The Orchestrator selects which processors to run as a pipeline based on the metadata that accompanies the data. One approach is to agree with the metadata stakeholders that it represents domain elements that add semantic value to the data; a composable approach is to treat the metadata as a structural description of the data. Once the pipeline finishes, the Data Provider can receive its corresponding results.

\textit{As an example, consider a tax calculation microservice that interacts with the product catalogues of an international store. Each time a tax calculation is required, it might involve products from different categories and be situated in different country contexts. The metadata representing the product list structure is combined with the metadata identifying the country. The orchestrator triggers the processors specific to the combination of category and country in each product. Such processors might be local or remote. Finally, when all is run, the orchestrator provides an output with the total tax value, and the Flexible Ingestor returns it as a response to the Data Provider that started the request.}

\begin{figure}
  \centering
  \includegraphics[width=.75\linewidth]{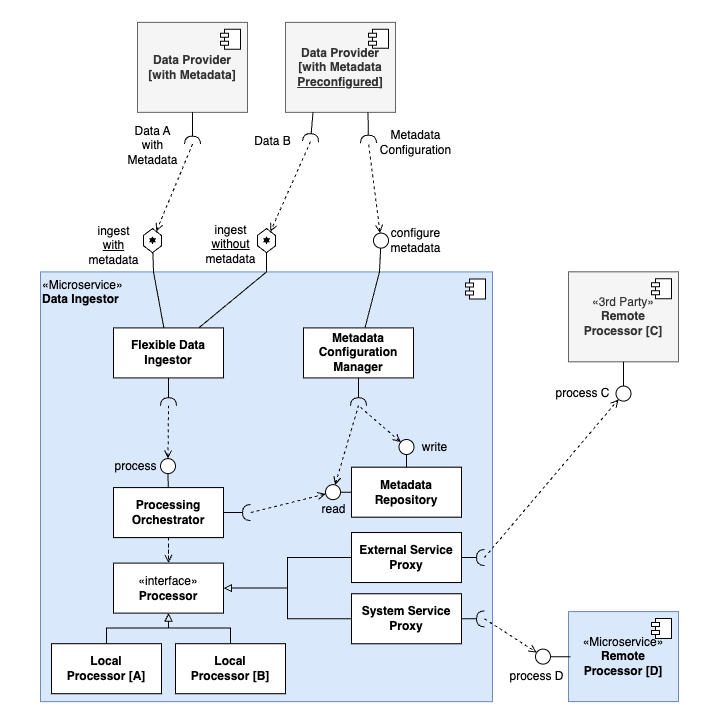}
  \caption{The components that implement the responsibilities of Ingestion and Processing of this concern}
  \label{fig:ing-proc-1}
\end{figure}

The processors can do more than process a piece of data with its metadata. One particular use case is when the ingestion is requested not for an immediate calculation, but to aggregate or consolidate the results for a later query. In this case, one processor serves as the Result Sink, writing the results to a Processed Data Repository. There are two ways to query the results: a flexible read, which returns the data as it is accompanied by descriptive metadata; and a normalised read, which provides a normal structure, with metadata used to normalise the record. The Results Server offers both approaches, fetching the existing metadata from the Metadata Repository and relying on the Normalizer to transform.

\textit{Here, an illustration of this scenario. Consider a smart city platform that ingests data from a variety of providers, such as the municipality informing road closures for renovations and the local university announcing lectures open to the community. For each ingestion, the data provider does not expect a synchronous result. But aggregating the events of interest for a given period might be interesting. In this case, the Sink stores the processed data if any processing occurs. Clients of the platform can query for the normalized list of events happening in a given week.}

\begin{figure}
  \centering
  \includegraphics[width=0.75\linewidth]{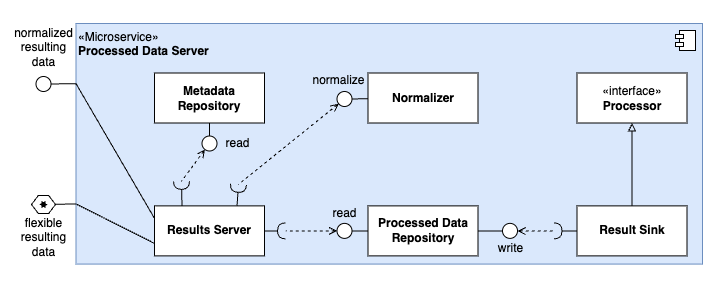}
  \caption{The components that implement the responsibility of Serving aggregated results of this concern}
  \label{fig:ing-proc-2}
\end{figure}

\subsection*{Concern 2: Moderation}

Not all the Data Providers that interact with the Ingestor are able or willing to provide metadata. For example, the provider is a public API consumed by multiple other systems. The lack of metadata support challenges the integration with the Ingestor.

Two other types of Data Providers do not support metadata:
\begin{enumerate}
  \item the Passive Data Provider, the one that offers its data upon request, and
  \item the Active Data Provider, that autonomously sends its data to interested parties.
\end{enumerate}

To address the lack of metadata essential to the  Ingestor, it is necessary to enrich the data on the Providers’ behalf. For that, there is the Metadata Enricher, a microservice that intermediates communication between a given Provider and the Ingestor. The Enricher component adds the specific metadata and forwards the data to the Ingestor. The “enrich” operation is publicly offered, so Active Providers can use it. For the Passive Providers, the microservice includes a Fetcher component that requests the data and passes it to the “enrich” operation.

The Metadata Enricher needs to be dedicated to a specific type of Data Provider so it can define the appropriate metadata.

\textit{For instance, consider a platform for personal organization that wants to integrate with a weather and forecast public API. The platform has a flexible ingestor as depicted in the previous section. The weather API is a Passive Data Provider for the platform’s ingestor. Then, the platform created a dedicated weather-and-forecast metadata enricher, based on the documentation and specification of the API.}

\begin{figure}
  \centering
  \includegraphics[width=0.4\linewidth]{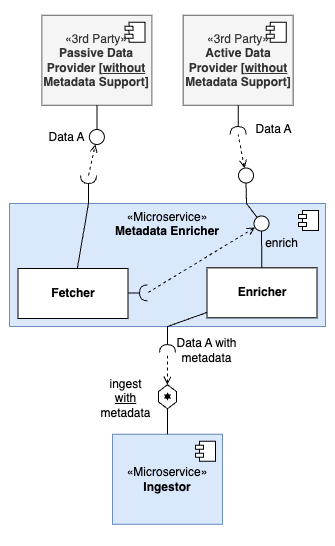}
  \caption{The Metadata Enricher in the application context of this concern}
  \label{fig:mod-1}
\end{figure}

Flexible ingestion might not always be the case, which challenges the integration: when a Service Provider offers an operation of interest to a Service Consumer, but neither the Provider has a flexible ingestion, nor the Consumer can change to either adapt its data structure or add metadata. To overcome that, there is the Metadata-Driven Moderator, which exposes a flexible input operation via the Transformation Trigger component, triggering the processing of the data in the Consumer shape into the Provider’s. This transformation happens based on the metadata descriptors stored in the Metadata Repository. Since it depends only on the metadata, not on the specific data structures, the same Moderator can integrate different Consumers with different Providers, as long as the Moderator has the appropriate metadata descriptors. Additionally, the Moderator can operate in various ratios, not only one-to-one Consumer and Provider.

Data structure transformation occurs under the Processing Orchestrator, as described in the previous section.

\textit{Consider the following scenario as an illustration. A marketplace has a legacy API for adding new stores; a digital store has its own e-commerce but is seeking to expand its business by integrating its store into the marketplace. The API provides the registration of stores with a fixed structure, whereas e-commerce has a behavior solely designed by the interaction with its own frontend. The integration strategy was the marketplace providing the Moderator.}

\begin{figure}
  \centering
  \includegraphics[width=0.5\linewidth]{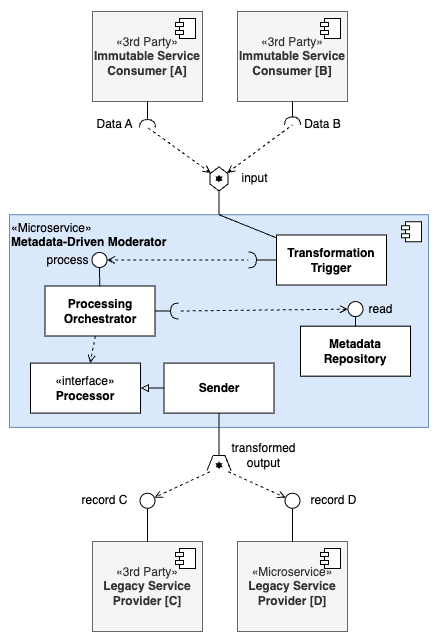}
  \caption{The Metadata-Driven Moderator}
  \label{fig:mod-2}
\end{figure}

\subsection*{Concern 3: External Consumption}

\begin{figure}
  \centering
  \includegraphics[width=0.75\linewidth]{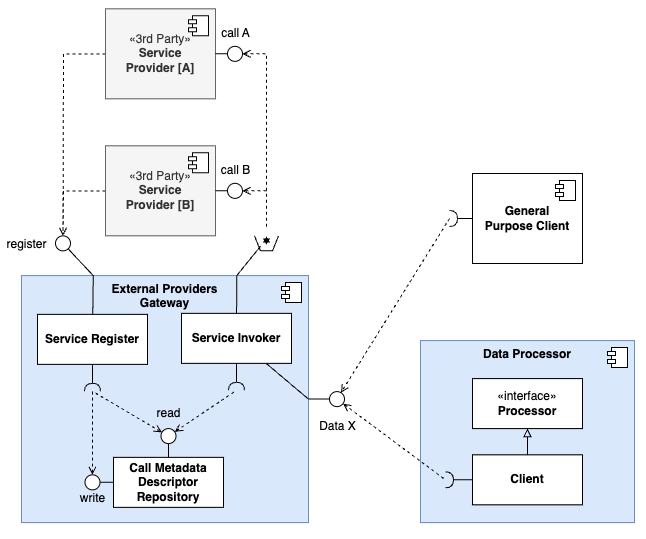}
  \caption{The External Consumption concern}
  \label{fig:ext-cons}
\end{figure}

It is common to rely on third-party services to enhance the functionality and extend the capabilities of an application. These external services, such as \textsc{Service Provider [A]} and \textsc{Service Provider [B]} (from Fig. \ref{fig:ext-cons}) provided by various vendors, offer specialized features that can be seamlessly integrated into our system. However, one of the primary challenges is managing the diverse data formats and protocols used by different service providers. Furthermore, each provider operates within a distinct domain; but for our application, despite these differences, these service providers can be seen as similar in the sense that they all contribute to the overall functionality.

The External Providers Gateway ensures seamless integration of external services into the system and comprises several key components. The Service Register is responsible for registering external services, making them available for consumption. It reads and writes to the Call Metadata Descriptor Repository to maintain a repository of available services and their associated metadata.

The Service Invoker is responsible for invoking external services based on metadata stored in the repository. It acts as the execution engine, translating metadata into actionable service calls and managing the communication with external services.

With these roles covered, the Gateway dynamically accommodates new Service Providers while providing an abstraction for its clients. Particularly, a Gateway’s client is when a Data Processor acts similarly to the Proxy case mentioned before: in a data processing pipeline, there is a step of collecting extra data for enrichment or crossing.

\textit{Consider an application that, among other details, enriches its domain data with both forecast and traffic information. In this scenario, there are two service providers, one for the forecast and the other for traffic information. The Service Register component registers them both by storing their associated metadata in the repository. When the application needs enrichment, one Processor requests data from the Gateway, which handles it internally with the Service Invoker. Note that, despite the distinct domains of weather forecasting and traffic information, these service providers are perceived as similar for our application, as they both provide data for enrichment.}

\section{Results of the Scenario-based Method}

In this section, we explore the proposed RA in a conceptual approach, considering three scenarios designed to showcase the FRs and QAs. In each, we discuss how the RA's components resolve the scenario's stimuli. Furthermore, we illustrate each resolution with a diagram in which we highlight in blue the RA's components and, as a pink star, the elements impacted by change.

\subsection{Metadata-based Platform Onboarding New Providers (Scenario 1)}

\begin{figure}
    \centering
    \includegraphics[width=0.5\linewidth]{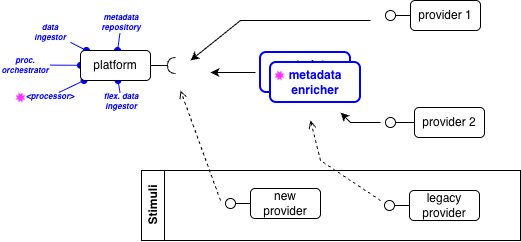}
    \caption{Scenario 1 resolved with RA's components; in blue, the RA's components; in pink, the required changes}
    \label{fig:scen1-resolved}
\end{figure}

In scenario 1, the \textsc{platform} receives heterogeneous data from \textsc{providers 1} and \textsc{2}, and this is further challenged by adding a \textsc{new provider} and a \textsc{legacy provider}. This scenario addresses \textbf{FR-1}, \textbf{FR-2}, \textbf{FR-3}, \textbf{FR-4}, \textbf{QA-1}, \textbf{QA-2}, and \textbf{QA-4}.

As illustrated in Fig. \ref{fig:scen1-resolved}, we present the scenario with the RA's components that are pertinent to this situation, highlighted in blue. The stacked \textsc{metadata enricher}s are there to overcome providers' non-compliance with including the required metadata; the \textsc{platform} is connected to a subset of the RA's internal components to ensure the flexibility of its API, as well as to act accordingly when processing the heterogeneous, metadata-enriched data it receives through such an API.

To handle the stimuli of this scenario, the changes required are around the elements highlighted with the pink stars in Fig. \ref{fig:scen1-resolved}. When adding a new provider, two aspects must be considered: its compliance with metadata use and the information the metadata contains. Regarding the former, when the new provider complies, then \textbf{NoC} no change is required; otherwise, an \textbf{ESA} external subsystem needs to be added -- the enricher. When it comes to the latter, the metadata from the new provider can be something already handled by the platform, in which case \textbf{NoC} no change is required; but if there is new information and processing to be made, then a \textbf{pCA} pluggable class is added, a new concrete implementation of the \textsc{<processor>} interface, to be mapped to these new metadata and to be plugged into the \textsc{processing orchestrator}.

Then, we indicate how this scenario and the RA components address each FR.

\begin{description}
    \item[FR-1 - heterogeneous data ingestion] the role of the \textsc{flexible data ingestor} is to define a flexible API contract, such that heterogeneous data can be passed;

    \item[FR-2 - contextual data characterization] on the providers' side, they either send the metadata along as characterization, or the \textsc{metadata enricher}s act on their behalf; internally, the \textsc{metadata repository} represents the dedicated handling of these data pieces;

    \item[FR-3 - data-driven processing adaptation] the \textsc{processor orchestrator} dynamically employs concrete \textsc{<processor>}s according to the metadata accompanying the data;

    \item[FR-4 - provider-side characterization bridging] when the providers do not comply with the use of metadata, the \textsc{metadata enricher} bridges over this gap;
\end{description}

With the resolution of this scenario, it is possible to see that the RA enables the same service -- the platform -- to be reused by different and heterogeneous clients -- the providers --, which is the \textbf{QA-1 - service reusability across structural heterogeneity}. It also demonstrates the \textbf{seamless extensibility of providers} and \textbf{processors}, which are respectively \textbf{QA-2} and \textbf{QA-4}.

\subsection{Service Connecting Heterogeneous Consumers and Providers (Scenario 2)}

\begin{figure}
    \centering
    \includegraphics[width=0.75\linewidth]{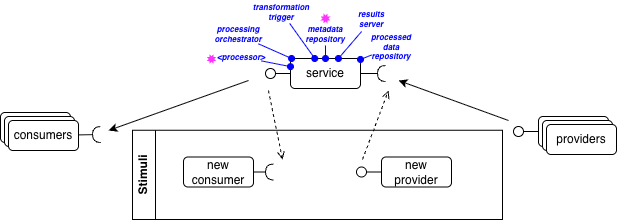}
    \caption{Scenario 2 resolved with RA's components; in blue, the RA's components; in pink, the required changes}
    \label{fig:scen2-resolved}
\end{figure}

In scenario 2, the \textsc{Service} bridges a set of \textsc{providers} and \textsc{consumers} that otherwise would not be directly possible. That is stimulated by adding a \textsc{new provider} offering new capabilities and adding a \textsc{new consumer} presenting new needs. It addresses \textbf{FR-3}, \textbf{FR-5}, \textbf{FR-6}, \textbf{FR-7}, \textbf{QA-1}, \textbf{QA-2}, and \textbf{QA-3}.

Fig. \ref{fig:scen2-resolved} presents the accommodation of the stimuli in the scenario using the RA's components, following the same schema as Fig. \ref{fig:scen1-resolved}. The internal components of the \textsc{service} are responsible for representing, using metadata, the transformations between consumers and providers, triggering processing for a specific consumer request, and serving the final results.

The changes required in this application of the RA are related to the elements highlighted with the pink stars in Fig. \ref{fig:scen2-resolved}. To add a new consumer or provider, the required changes are similar: the metadata repository needs to reflect the new participants, i.e., a new target shape for consumers or a new source for providers, which is a \textbf{CfC} configuration change; then, it also needs a new processor to be plugged in to the processing orchestrator, a \textbf{pCA} that makes it possible to implement the needed algorithms for the new steps in transforming the data.

Each FR is addressed as follows:

\begin{description}
    \item[FR-3 - data-driven processing adaptation] the \textsc{processing orchestrator} dynamically employs concrete \textsc{<processor>}s according to the metadata describing the transformations required between providers and consumers;

    \item[FR-5 - structural mediation between incompatible parties] the \textsc{service} acts as the RA's \textsc{metadata-driven moderator} providing to the \textsc{consumers} needs, according to the concrete capabilities of the \textsc{providers};

    \item[FR-6 - dynamic external service integration] based on the content of the \textsc{metadata repository}, the \textsc{service}'s code can be generic and work for a \textsc{provider} defined during runtime;

    \item[FR-7 - result aggregation and flexible retrieval] the \textsc{metadata repository} can describe complex transformations that involve sourcing multiple \textsc{providers} for a single \textsc{consumer}'s need. 
\end{description}

Applying the RA to the resolution of this scenario achieves \textbf{QA-1} in a more nuanced way: the \textsc{service} can be reused by its external clients \textsc{consumer} -- as in the previous scenario -- as well as by the \textsc{providers} it abstracts in its API. It is possible to see that this resolution with the RA also supports \textbf{QA-2} and \textbf{QA-3} by updating the metadata configuration and eventually plugging in new processors.

\subsection{Enriching Capabilities with $3^{rd}$-party (Scenario 3)}

\begin{figure}
    \centering
    \includegraphics[width=0.75\linewidth]{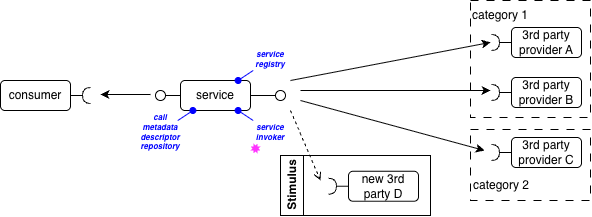}
    \caption{Scenario 3 resolved with RA's components; in blue, the RA's components; in pink, the required changes}
    \label{fig:scen3-resolved}
\end{figure}

In Scenario 3, \textsc{service} enhances its capabilities with different \textsc{$3^{rd}$-party providers A, B} and \textsc{C}. This context is being stimulated by a \textsc{new $3^{rd}$-party D}. It addresses \textbf{FR-2}, \textbf{FR-5}, \textbf{FR-6}, \textbf{FR-7}, \textbf{QA-1}, \textbf{QA-2}, and \textbf{QA-4}.

In blue in Fig. \ref{fig:scen3-resolved}, there are the internal components of \textsc{service} that belong to the RA. They support a seamless integration of various $3^{rd}$-party providers by registering metadata describing the technical details of calling them.

This scenario's stimuli are handled by the \textsc{service invoker} implementing the protocol (e.g., a REST service caller or a gRPC invoker), which in various cases is a \textbf{NoC} no change. Instead, when a new external provider registers itself with a new protocol, then a new type of \textsc{service invoker} is required to be plugged in, i.e., a \textbf{pCA}.

The requirements are implemented as follows:

\begin{description}
    \item[FR-2 - contextual data characterization] the \textsc{call metadata descriptor repository} stores metadata about the protocols and technical details for calling the $3^{rd}$-party services;

    \item[FR-5 - structural mediation between incompatible parties] the \textsc{service} allows its \textsc{consumer}s to indirectly integrate with a range of \textsc{$3^{rd}$-party providers}, even if that is not the main goal;

    \item[FR-6 - dynamic external service integration] the mechanics involving \textsc{service registry}, \textsc{call metadata descriptor repository}, and \textsc{service invoker} allows to virtually any \textsc{$3^{rd}$-party provider} to be integrated;

    \item[FR-7 - result aggregation and flexible retrieval] especially but not restricted to when multiple \textsc{$3^{rd}$-party provider}s belong to different categories, the \textsc{service} can potentially aggregate their data as part of its capability.
\end{description}

The dynamic registration and call of external services meet the \textbf{QA-1}. In this scenario, we can consider an overlap of the roles "provider" and "processor" for each $3^{rd}$-party provider: as service providers, each can fall at different points on the spectrum from data provider to data processor. With that, it meets \textbf{QA-2} and \textbf{QA-4} as well.

\subsection{Summary}

In summary, we present Table \ref{tab:scen-results-summary} with the stimuli of each scenario, and the respective observed level of impact (according to the taxonomy presented in Subsec. \ref{subsec:scenarios}).

For scenario 1, there are two steps for the impacts: the first depends directly on the stimuli, and the second depends on the novelty of the metadata. For scenario 2, regardless of the stimuli the impact is the same; it only depends whether new transformations are required. Finally, for scenario 3, the only impact depends whether there is a processor for the communication protocol.

\begin{table}[]
\centering
\begin{tabular}{c|l|cc}
\multicolumn{1}{l}{\textbf{Scenarios}} & \textbf{Stimuli}      & \multicolumn{2}{l}{\textbf{Impacts}}                             \\\hline
\multirow{2}{*}{1}                     & new provider          & \textbf{NoC}                  & (know meta) \textbf{NoC}                           \\
                                       & legacy provider       & \textbf{ESA}                  & (new meta) \textbf{pCA}                            \\\hline
\multirow{2}{*}{2}                     & new provider          & \multirow{2}{*}{\textbf{CfC}} & \multirow{2}{*}{(new transformation) \textbf{pCA}} \\
                                       & new consumer          &                      &                                           \\\hline
3                                      & new external provider & \multicolumn{2}{c}{\textbf{NoC}; unless new protocol, then \textbf{pCA}}           
\end{tabular}
\caption{Summary of the types of impact for each scenario's stimuli}
\label{tab:scen-results-summary}
\end{table}

\section{Results of the Case Study Method}\label{sec:results-cases}

In this section, we present the four cases included in the study design, with the purpose of observing the RA performing with the scenarios applied to real systems. Each of the cases is also illustrated with a diagram, in which the green label indicates the scenario element, in purple the RA's components, and in orange the components of the case's system.

\subsection{Open Data Hub}

\begin{figure}
    \centering
    \includegraphics[height=4cm]{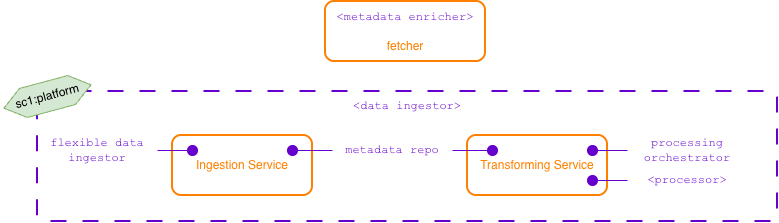}
    \caption{Representation of the Open Data Hub case, tagged with the scenarios' elements, and mapped into the RA's components}
    \label{fig:case-odh}
\end{figure}

The case of Open Data Hub (illustrated in Fig. \ref{fig:case-odh}) matched scenario 1 entirely, and it also matched a particularity of scenario 2. Despite our mapping of requirements to scenarios, this case was strongly aligned with \textbf{FR-7 - result aggregation and flexible retrieval}, as it is the means to achieving their goal. In other words, the purpose of their platform was to collect data from various sources and aggregate it into a single, structured API to facilitate its integrated use. Nonetheless, their structural features were strongly aligned with our designed scenario 1.

The first important aspect to mention about this case, regarding scenario 1, is how the RA's \textsc{data ingestor} was implemented: they decided to separate the concerns of ingestion and processing into two independent microservices, allowing custom scalability and deployment configurations for each. For instance, they valued availability for their \textbf{Ingestion Service}, while they could favor scheduling and load balancing for their \textbf{Transforming Service}.

Another interesting feature of this case was the driver for implementing the RA's \textsc{metadata enricher}: initially, they struggled with passive data providers, i.e., external systems that would only provide data upon request. That is why their component is called \textbf{Fetcher}, which later they realised could be the type of component to handle enrichments when needed.

In this case, the stimuli of scenario 1 map into the following concrete situations: 

\begin{itemize}
    \item a new provider describing its own metadata happens when a project starts a partnership with the Open Data Hub, and agrees on sending their data autonomously, duly accompanied with the metadata descriptors -- one instance would be a municipality that integrates their own system with the project;\\
    In this situation, there was no need for an enricher, but it did require a new set of processors because the metadata added information that required new algorithms to run.

    \item a new legacy provider would map into, for example, an API that provides an access key to their resources, but they are passive and do not add their own metadata -- for example, a regional forecast API.
    In this situation, a fetcher-and-enricher was needed to gather the data and add metadata before sending it to ingestion, but no new processors were needed, as there were already processors for meteorological data.
\end{itemize}

For scenario 2, and more specifically \textbf{FR-7}, it falls under the responsibilities of the \textbf{Transforming Service}. The raw data, along with its respective metadata, are used as input for the transformations into the normal format adopted by their platform. This format structures in a common way a wide variety of data records of different domains under higher abstractions; for example, a "station" represents a data generator and is applied either to a point-of-interest like a hotel or to a e-charging parking spot, or a "measurement" represents a data piece and is applied either to the vacancy status of a hotel room or to the temperature status given by a thermal sensor on the mountains. They also support transforming the data served, but in a simpler way than a full reshaping; instead, they allow filtering by attributes within a data record.

\subsection{Catch \& Solve}

\begin{figure}
    \centering
    \includegraphics[height=4cm]{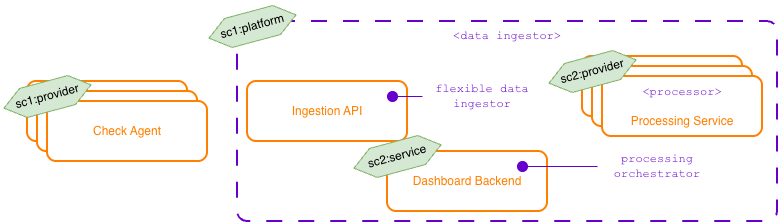}
    \caption{Representation of the Catch\&Solve case, tagged with the scenarios' elements, and mapped into the RA's components}
    \label{fig:case-cs}
\end{figure}

The case of the Catch\&Solve platform (represented by Fig. \ref{fig:case-cs}) partially matches scenarios 1 and 2, as it depends on the interpretation of the roles either for the system's components or the scenarios' elements. It matches scenario 1, considering the system's \textbf{Ingestion API} as an important piece of the scenario 1 Platform, given its role in ingesting data from data providers. At the same time, the interaction between the \textbf{Dashboard Backend} and the \textbf{Processing Services} can be expressed as the service consumption defined by the Platform and Providers in scenario 2.

In this case, we observed a technical decision regarding the implementation of the RA's \textsc{Data Ingestor} similar to the Open Data Hub case: the separation between ingestion and processing. In this case, the availability of the \textbf{Ingestion API} was the main driver, leading to separating it into a more compact piece of software to keep it replicated and online.

The stimuli of the scenarios are observed in the following concrete situations of this case:

\begin{itemize}
    \item a new provider (from scenario 1) happened when they decided to revisit an existing Check Agent to embed a new piece of information into its data;\\
    When it happened, only updates were required for the Check Agent and its respective processing service.

    \item a new service provider (from scenario 2) happened as a response to the first stimulus mentioned, when a processing service was updated; due to backward compatibility, the updated version was a different component released.
\end{itemize}

\subsection{Digi Dojo}

\begin{figure}
    \centering
    \includegraphics[height=4cm]{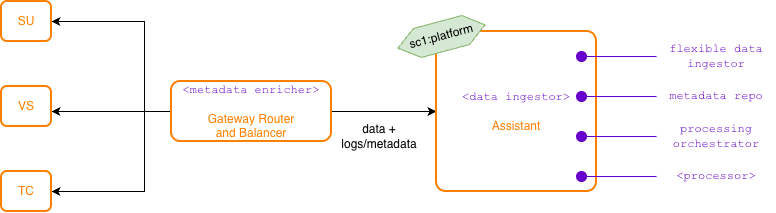}
    \caption{Representation of the Digi Dojo case, tagged with the scenarios' elements, and mapped into the RA's components}
    \label{fig:case-dd}
\end{figure}

Fig. \ref{fig:case-dd} illustrates the matching of the Digi Dojo case with scenario 1. In this case, it is possible to see an interesting implementation of the RA's component \textsc{Metadata Enricher} in the system's component \textbf{Gateway Router and Balancer}. In this application, it serves as a notifier for the \textbf{Assistant}, informing it of each new request -- that is, the data being ingested. The interesting aspect is the complete division this enricher creates, in which there is no need to directly integrate the domain microservices -- \textbf{Startups and Users}, \textbf{Virtual Spaces}, and \textbf{Tasks and Calendars} -- with the \textbf{Assistant}.

The stimulus from scenario 1 that would be visible in this case would occur with the evolution of the system, i.e., when new responsibilities are implemented in new microservices or when an existing one is split into multiple ones.

\subsection{Metrics Platform}

\begin{figure}
    \centering
    \includegraphics[height=2cm]{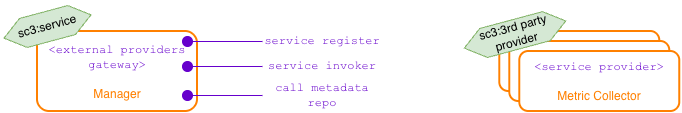}
    \caption{Representation of the Metrics Platform case, tagged with the scenarios' elements, and mapped into the RA's components}
    \label{fig:case-mp}
\end{figure}

This case was the only one that matched scenario 3. In Fig. \ref{fig:case-mp}, we present it visually, but the Consumer is omitted because there was no impact on the analysis. It is possible to see that the \textbf{Manager} implemented all the RA's components related to scenario 3, namely \textsc{Service Register}, \textsc{Service Invoker}, and \textsc{Call Metadata Descriptor Repository}. In this case, to provide its clients with a metrics report on their architecture, the \textbf{Manager} relies on the various \textbf{Metric Collectors}, which act as the Service Providers in the scenario.

An interesting highlight for this case regards the distribution of responsibilities. Similarly to the cases of Open Data Hub and Catch\&Solve, the maintainers of the Metrics Platform had concerns about drawing a hard boundary between two responsibilities -- here, registering and invoking $3^{rd}$-party services --, but in this case, the decision was the opposite of the previous cases. It was considered to implement each responsibility in a dedicated microservice, but this was discarded due to the associated costs of maintaining and deploying that infrastructure. Instead, they decided to package both in a single deployable unit to reduce these costs. Nonetheless, the intentions of splitting originated from the specific measures the invoking of services would require, such as security, scalability, and, in particular, the possibility to easily schedule re-runs (for this case's domain, some invocations are done in repetition, e.g., every 5 minutes or twice every day), instead of solely invoking on demand by the client.

The stimulus for scenario 3, when a new service provider is to be added, occurred when a new metric collector was added. In this project, the same maintainers of the \textbf{Manager} hold ownership of the \textbf{Metric Collector}s currently available -- but they implemented this architecture to enable seamless extension of the collectors. In this case, it happened when adding a new collector, and it only required the \textbf{Metric Collector} to register itself in the \textbf{Manager} during bootstrap at runtime.

\section{Discussion}

\subsection{Combined Results Analysis}\label{subsec:combined-results}

The triangulation of the results from both methods covered all functional requirements and quality attributes of the proposed RA. We summarize this in Table \ref{tab:res-per-case-thru-scenarios}, where it crosses the FRs and QAs with the cases through the scenarios, e.g., FR-3 (3rd row) is assessed in the "Catch\&Solve" (3rd column) case through scenarios 1 and 2. It is important to note that Table \ref{tab:res-per-case-thru-scenarios} is a join of Tables \ref{tab:reqs-per-scenario} and \ref{tab:scen-summary} on the scenarios.

\begin{table}
    \centering
    \begin{tabular}{l|cccc|c}
             & Open Data Hub & Catch\&Solve & Digi Dojo & Metrics Platform & \#scenarios / \#cases\\\hline\hline
        FR-1 &    1          &    1         &   1       &                  & 1 / 3 \\\hline
        FR-2 &    1          &    1         &   1       &   3              & 2 / 4 \\\hline
        FR-3 &    1          &    1,   2    &   1       &                  & 2 / 3 \\\hline
        FR-4 &    1          &    1         &   1       &                  & 1 / 3 \\\hline
        FR-5 &               &    2         &           &   3              & 2 / 2 \\\hline
        FR-6 &               &    2         &           &   3              & 2 / 2 \\\hline
        FR-7 &    2          &    2         &           &   3              & 2 / 3 \\\hline\hline
        QA-1 &    1          &    1,   2    &    1      &   3              & 3 / 4 \\\hline
        QA-2 &    1          &    1,   2    &    1      &   3              & 3 / 4 \\\hline
        QA-3 &               &    2         &           &                  & 1 / 1 \\\hline
        QA-4 &    1          &    1         &    1      &   3              & 2 / 4 
    \end{tabular}
    \caption{FRs and QAs observed in each case studied, and the scenarios in which they appear}
    \label{tab:res-per-case-thru-scenarios}
\end{table}

The combined analysis makes explicit that \textbf{FR-2} \textit{contextual data characterization} is the prerequisite that enables \textbf{FR-3} through \textbf{FR-7}. The use of metadata as an architectural mechanism has a cascading effect: much of the RA's coverage can be traced back to this central design decision. This has practical implications for adoption, as even a partial instantiation of the RA (e.g., limited to metadata-driven processing) already unlocks a substantial portion of its benefits, with the remaining concerns composable on top as needs arise.

\finding{A}{Metadata is the central enabling mechanism of the RA.}{Metadata is the backbone of the solutions in the three concerns of the RA: in \textit{Data Ingestion and Processing}, with it the heterogeneity can be handled, and the dynamic processing can happen; in \textit{Moderation}, metadata is the enabler for connecting legacy providers to non-complying clients; and in \textit{External Consumption}, the metadata describing the service providers is the reason why no changes are required to any stimuli.}

From the evaluations, we could see that the \textbf{FR-1} \textit{heterogeneous data ingestion} was a key step in achieving the \textbf{QA-1} \textit{service reusability across structural heterogeneity}. The Open Data Hub case made it clear that integrating numerous new data providers into a single API operation was straightforward. Despite the heterogeneity among the providers, the platform could be reused seamlessly. On a similar note, \textbf{FR-1} also presents positive impacts on the actualization of \textbf{QA-2} \textit{seamless extensibility of providers}.

There was already evidence of this in the analysis of Scenario 1: no changes are required when a new data provider sends a known metadata type. Even when corner cases occur, the change impact is reduced, requiring either a pluggable implementation of the \textsc{Processor} interface for new types of metadata or a \textsc{Metadata Enricher} when the client does not comply with the metadata requirements.

The direct implementation of the \textbf{FR-1} nonetheless presents issues. From this stems the challenge of handling unforeseen data representations, which is addressed by the \textbf{FR-2} \textit{contextual data characterization}. The use of metadata -- the adopted approach to \textbf{FR-2} -- is the most characteristic feature of the proposed RA, enabling a considerable number of other RA's functional requirements and quality attributes. Metadata is at the heart of the solutions for \textbf{FR-2} throughout \textbf{FR-7}.

\textbf{FR-3} \textit{data-driven processing adaptation} was done in the proposed RA via the metadata from \textbf{FR-2}. A first key use of metadata is done by the \textsc{Processing Orchestrator} when dynamically employing concrete \textsc{Processor} implementations to process a data record. From the analysis of Scenario 1, the impacts of these changes are considered minimal, as it consists of the straightforward solution of adding a pluggable class (\textbf{pCA}) that enables either a local implementation or an adapter to a remote system.

We could also evaluate this in practice in all the cases studied. The Catch\&Solve system and the Digi Dojo had a 1-to-1 approach between a high-level abstraction (represented in the metadata) and the concrete \textsc{Processor}; on a different approach, the Open Data Hub had metadata indicating smaller, low-level information in the metadata, which led the \textsc{Processing Orchestrator} to compose a set of concrete \textsc{Processor}s. That provides evidence for \textbf{QA-4} \textit{seamless extensibility of processors}, and on a step further, of reusability of these processors in different contexts. But that comes at the expense of a slightly more complex \textsc{Processing Orchestrator} to dynamically match multiple metadata to multiple \textsc{Processor}s.

Both approaches are valid instantiations of the same RA components -- the \textsc{Processing Orchestrator} and the \textsc{Processor} interface -- and neither required deviating from the RA's structure. This variability in granularity, not prescribed by the RA, suggests that the RA is flexible enough to accommodate different levels of processing complexity. Nonetheless, it trades off the compositional approach, which provides greater reusability of individual processors across different contexts, for a more sophisticated orchestrator that dynamically matches multiple metadata values to multiple processor implementations.

\finding{B}{The RA accommodates heterogeneous processor granularity.}{The solution involving the \textsc{Processing Manager} and \textsc{<Processor>}s implementations mapping to metadata is flexible to support from coarse-grained mapping between metadata and implementation, as seen in the case of Catch\&Solve, to a fine-grained, composable one as seen in the Open Data Hub. This flexibility makes the RA adaptable, where the same components are applicable to different scenarios.}

Imposing metadata as part of the contract between a service and its clients causes friction, as we saw with the Open Data Hub, which had some non-compliant data providers, and with Digi Dojo, where it was desirable to have domain services decoupled from the Assistant. The proposed RA encompasses components to address and mitigate these issues, providing \textbf{FR-4} \textit{provider-side characterization bridging} and \textbf{FR-5} \textit{structural mediation between incompatible parties}. In Scenarios 1 and 2, the results provided evidence of the controlled impact of these intermediate components -- the \textsc{Metadata Enricher} and \textsc{Metadata-Driven Moderator}.

Across all three scenarios, the only stimulus that requires a more impactful change (an \textbf{ESA}) is the onboarding of a legacy provider that cannot supply metadata. All other stimuli are more frequent and resolve at \textbf{NoC}, \textbf{CfC}, or \textbf{pCA}, which are of lower impact. This means the RA confines architectural growth to well-justified cases: a new deployable unit is required only when a client is structurally incapable of participating in the metadata contract. The cost is proportional to the degree of incompatibility, not to the act of extending the system.


In the Digi Dojo and Open Data Hub cases, we saw examples of the \textsc{Enricher} as the solution. Implementing both requirements led to \textbf{QA-2} and \textbf{QA-3}, providing \textit{seamless extensibility for consumers}. Also, the Digi Dojo case highlighted the \textsc{Enricher}'s impact on the dependency between components, as it enabled complete decoupling of the domain services from the Assistant.

The \textsc{Metadata Enricher} was designed primarily to bridge non-compliant clients (\textbf{FR-4}), enabling legacy or unwilling providers to participate in the metadata contract without modifying their own implementation. The Digi Dojo case, however, revealed a secondary benefit: the component also functions as a structural decoupling mechanism between otherwise unrelated domain services. By routing domain microservice traffic through the Gateway Router and Balancer before it reaches the Assistant, the system achieved a complete independence between its domain services and the research support component -- a separation that was architecturally desirable but not the original motivation for the enricher. This emergent benefit suggests the component's value extends beyond the reusability dimension it was explicitly designed to address.

\finding{C}{The Metadata Enricher may carry decoupling as a secondary benefit.}{The \textsc{Metadata Enricher} is structurally influenced by the \textsc{Decorator} design pattern from Object-Oriented paradigm \citep{Gamma1995DesignSoftware}, which provides a decoupled way to add behavior and, as seen in the Digi Dojo case, architectural separation of concerns.}

The taxonomy analysis across all three scenarios shows that the RA consistently avoids the two most disruptive change types: source code change and API contract change. The predominant impacts are \textbf{NoC}, \textbf{CfC}, and \textbf{pCA}. This is mapped to concrete situations in the cases (Sec. \ref{sec:results-cases}). In Catch\&Solve, a new provider triggered only updates to the Check Agent and its processing service; in Metrics Platform, a new collector required only runtime self-registration. These observations characterize \textit{how} the RA bounds service proliferation: the cost of extending the system is proportional to the novelty introduced, not to the act of extension itself.

\finding{D}{Change impact with the RA is bounded and predictable.}{The most impactful change observed is the \textbf{ESA}, which is an impact that only happens in corner cases (e.g., onboarding a provider that does not comply to the use of metadata), and furthermore, is still a mild impact considering the adopted taxonomy. In other words, implementing the RA leads towards a more seamless evolution of the architecture.}

There is a close relationship between \textbf{FR-6} \textit{dynamic external service integration} and \textbf{FR-7} \textit{result aggregation and flexible retrieval}. In the Metrics Platform case, these requirements were implemented in collaborating components: concrete metric collectors could register with the Manager, and their metrics could be aggregated into the client's result. These two FRs are an important asset for the extensibility provided by the proposed RA, as they support embedding new external capabilities into the system and enabling new consumer interactions as clients. That translates into support for the quality attributes \textbf{QA-2} and \textbf{QA-3}.

Both Open Data Hub and Catch\&Solve independently arrived at splitting the \textsc{Data Ingestor} into two separate deployable units, each driven by distinct operational concerns; Open Data Hub prioritized availability for ingestion while favoring scheduling and load balancing for processing, while Catch\&Solve separated the two for similar availability reasons, keeping the ingestion API compact and replicated. These decisions were made independently, without prescriptions from the RA, yet they consistently align with the RA's component boundary between ingestion and processing. The recurrence across independent systems suggests that this boundary is structurally well-placed, coinciding with a natural operational fault line that practitioners discover on their own.

\finding{E}{Ingestion and processing separation recurs as an independently motivated decision.}{The use of metadata as approach to service reusability yields an important decoupling between ingestion and data processing; despite being sequential, the proposed RA accommodates independent operation and runtime for these tasks, as demonstrated in the Catch\&Solve and Open Data Hub cases.}

In conclusion, as illustrated in Table~\ref{tab:res-per-case-thru-scenarios}, the combined methods of scenario-based evaluation and case studies enabled the analysis of all seven functional requirements and four quality attributes of the proposed RA. The final analysis brought enough elements to sustain the proposed RA regarding the fulfillment of its design requirements, which ultimately answers \textbf{RQ2}: the adoption of the proposed RA is an effective architectural artifact in providing or improving the reusability of a service, favoring the evolution through a flexible API contract that requires no changes over time, as well as through a well-established metadata-based approach, whose benefits span from data description to dynamic processing.

\textbf{QA-3} \textit{seamless extensibility of consumers} was observed in only one case (Catch\&Solve, through scenario 2), as reflected in Table~\ref{tab:res-per-case-thru-scenarios}. While the RA's design supports consumer extensibility through the \textsc{Metadata-Driven Moderator}, and the scenario-based analysis confirms the mechanism conceptually, the empirical grounding for this quality attribute is thinner than for the remaining six. This does not invalidate the claim, but it does calibrate the confidence with which it can be asserted. 

\finding{F}{The proposed RA fulfills all 7 FRs and all 4 QAs}{With the combination of scenario-based and the case study methods, we found enough evidence to support all the RA's requirements; still, the conditions covered focused on features that were specialization of "consumer", such as providing data or other services (expressed in QA-2), hence QA-3 has a concrete yet thinner support than the others.}

\subsection{Related Works}

The evolution of systems and their architecture is subject to challenges, often documented as anti-patterns or bad smells. Anti-patterns are recurrently adopted solutions that lead to negative consequences to the software \citep{brown1998antipatterns}, whereas bad smells are symptoms of poor design choices that erode the qualities of a system or architecture \citep{Fowler1999RefactoringCode}. SbA are also affected by these challenges \citep{Taibi2019MicroservicesTaxonomy,cerny2023catalog}, of which we highlight three:

\begin{itemize}
    \item \textit{Microservice Greedy} is reported as the explosion in the number of services in a system \citep{Mumtaz2021ADetection}, which is the architectural issue motivating this work;

    \item \textit{Duplicated Service} is defined as a set of similar services \citep{Mumtaz2021ADetection,Bogner2019TowardsSmells,sabir2019systematic}; this bad smell might stem from \textit{Greedy} and can be addressed by the pattern \textsc{Flexibilize the Ingestion} implemented by the \textsc{Flexible Data Ingestor} of this RA; and

    \item \textit{Co-Change Coupling} is said to be present when one service requires a change to reflect another service's change \citep{Mumtaz2021ADetection}; this bad smell indicates a tight coupling between two services, which might be due to a poor API design with rigid data structures. This can also be addressed by the proposed RA.
\end{itemize}

Reusability and modifiability are related design quality attributes, as both influence the evolution of an architecture \citep{Bass1997SoftwarePractice}. \citet{cobaleda2016reference} propose a RA to ease modifiability in personalised web applications. Their RA is centered on the \textsc{Personalisation Controller} module, which identifies and executes the appropriate personalisation. The personalisations are implemented as a set of components specialized in a goal or strategy. To isolate personalisation data from transactional domain data, they propose a separate module: the \textsc{Personalisation Model Administrator}, in which relevant data is replicated from the domain to support inference of contextual data that guides and enriches concrete personalisations.

The interaction established between \textsc{Personalisation Controller} and the specialized components in their work aiming at personalisation, is structurally very similar to our proposed RA's \textsc{Processing Orchestrator} and \textsc{<Processor>}s, aiming at supporting \textbf{FR-3} \textit{data-driven processing adaptation}. These structures reflect the \textit{Dependency Inversion Principle} of Object-Oriented Programming \citep{Martin2012TheArchitecture}, and can be considered implementations of the \textsc{Strategy} design pattern \citep{Gamma1997DesignSoftware}.

Furthermore, both approaches rely on metadata for employing concrete processing or customizations. But here there is a difference in how the metadata ownership is assigned: in their approach, the metadata (or \textit{inferred data}, as they refer to it) has an implicit nature, and the client does not own it; meanwhile, our RA relies on explicit metadata and is flexible regarding its provenance.

\subsection{Threats to Validity}

The evaluation methodology combines scenario-based assessment and case study research, each carrying its own validity threats, some of which are inherited from their respective methodological traditions and others that arise from the specific way they were applied in this work.

\textbf{Construct Validity}: The functional requirements and quality attributes defined in Subsec. \ref{subsec:refarch-design} were distilled directly from RQ1 by the authors. The requirements may not fully capture what service reusability demands in practice. A partial mitigation is that the RA was grounded in a pre-existing pattern language with aligned goals \citep{LinoDaniel2024PatternAPIs, daniel2025incrementing}, whose patterns were independently published and peer-reviewed, lending some external grounding to the reality .

The change impact analysis taxonomy (NoC, CfC, pCA, ESA, CSC, ACC) was created by the authors for this study. Nonetheless, to mitigate its negative impacts, it was an adaptation of what was demonstrated by \citet*{Zhao2002ChangeEvolution}, when the method was proposed.

\textbf{Internal Validity}: The scenarios were constructed by the same team that designed the RA, creating a risk that they were tailored to showcase the architecture's strengths rather than probe its limits. We attempted to mitigate it by systematically deriving the scenarios and keeping traces from the FRs and QAs prior to the analysis (Table \ref{tab:reqs-per-scenario}), and the stimuli were defined before the RA's components were applied to resolve them.

Furthermore, the case study relied on the authors' own access to and knowledge of the selected systems, three of which have direct institutional proximity to the research team. This proximity facilitates depth of observation but threatens independence of analysis. A mitigation present in the study is that maintainers of the systems were made available for clarifications, introducing at least one external perspective into the evidence gathering. Nevertheless, no formal protocol for interviews or data collection is reported, which limits the auditability of how case evidence was gathered and interpreted \citep{Runeson2009GuidelinesEngineering}.

The triangulation of two independent evaluation methods -- scenario-based assessment and case study -- serves as a structural mitigation for internal validity overall: conclusions that converge across both methods are less likely to be artifacts of either method's specific weaknesses.

\textbf{External Validity}: The four cases selected are all relatively small-scale systems. This concentration limits the generalizability of the findings to large-scale industrial systems, systems operating under strict regulatory constraints, or systems developed in different organizational cultures. No explicit mitigation for this threat is present in the study; broadening the case selection in future work to include larger, independently developed systems would strengthen this dimension.

The scenario-based method, as noted by \citet*{kazman2002scenario}, evaluates an architecture against a finite set of representative situations and cannot provide exhaustive coverage. Conclusions from the scenario analysis should therefore be understood as evidence of fitness-for-purpose in the described contexts, not as a general guarantee of architectural quality.

A mitigation for external validity is the deliberate use of realistic elements in each scenario's contextual description, as well as the subsequent grounding of those scenarios in actually existing systems through the case study. This connection between the abstract scenarios and concrete implementations increases the ecological validity of the evaluation beyond what a purely analytical assessment would offer.

\textbf{Conclusion Validity}: The triangulation of results across two methods (Subsec. \ref{subsec:combined-results}) strengthens conclusion validity by requiring that each FR and QA be evidenced both analytically, through change impact analysis, and empirically, through case observation. However, not all FRs and QAs were observed across all four cases (Table \ref{tab:res-per-case-thru-scenarios}); in particular, QA-3 was observed in only one case. This uneven coverage means confidence in conclusions varies across requirements, and claims about QA-3 rest on thinner empirical support than the remaining attributes.

An additional potential mitigation not explicitly discussed in the paper would be member checking: sharing the case study findings with the maintainers of each system for confirmation that the mapping between their implementation and the RA's components was accurately interpreted. If this was performed informally through the clarification exchanges mentioned in the study, making it explicit would strengthen the auditability of the conclusions.
\section{Conclusion}

This work addressed the challenge of service reusability in service-based architectures through an architectural design approach. Motivated by the recurring problem of services with replicated or similar functionalities proliferating across a system, we proposed a reference architecture whose central mechanism is the use of metadata to guide the ingestion, processing, and mediation of heterogeneous data.

The RA was designed following the pattern-based method for creating reference architectures \citep{Guerra2015RelatingArchitectures}, grounded in a pattern language specifically developed for metadata-driven service reusability \citep{LinoDaniel2024PatternAPIs, daniel2025incrementing}. This grounding ensured that each architectural component carries a well-motivated solution to a recurring problem, rather than being an ad hoc design decision. The architecture covers three concerns, each addressing a distinct dimension of the reusability challenge: Data Ingestion and Processing, Moderation, and External Consumption.

To answer RQ1, the proposed RA defines a structure in which a single service can absorb heterogeneous providers and consumers without requiring structural changes to its core components. This is achieved through a flexible ingestion contract, metadata-driven processing orchestration, and intermediary components -- the \textsc{Metadata Enricher} and the \textsc{Metadata-Driven Moderator} -- that bridge structural incompatibilities between parties that cannot or will not adapt to each other. The \textsc{External Providers Gateway} extends this principle to $3^{rd}$-party service integration, enabling dynamic registration and invocation of external capabilities without coupling the consuming components to any specific provider.

To answer RQ2, the RA was evaluated through a combination of scenario-based assessment and case study. The scenario-based method applied change impact analysis \citep{Zhao2002ChangeEvolution} to three representative scenarios, covering all seven functional requirements and four quality attributes defined for the RA. The results demonstrated that the most common type of change required when the architecture evolves is either no change at all, a configuration change, or the addition of a pluggable class, consistently avoiding the more disruptive source code and API contract changes. The case study method grounded these findings in four existing systems, each of which independently converged on architectural solutions that align with the proposed RA: Open Data Hub, Catch\&Solve, Digi Dojo, and Metrics Platform. The triangulation of both methods covered all defined requirements and attributes, providing converging evidence that the proposed RA is an effective architectural artifact for promoting service reusability.

Beyond the structural contributions, the case studies revealed nuances in how the RA's components can be instantiated in practice. The separation of ingestion and processing concerns into independent deployable units, observed in both Open Data Hub and Catch\&Solve, illustrates how the RA accommodates operational concerns such as availability and scalability without requiring modifications to its design. The Digi Dojo case demonstrated a non-obvious benefit of the \textsc{Metadata Enricher}: beyond its primary role of bridging metadata non-compliance, it can serve as a decoupling mechanism between otherwise unrelated domain services. These observations suggest that the value of the RA extends beyond the reusability dimension it was explicitly designed for.

As future work, first, the evaluation was conducted on relatively small-scale systems. Hence, applying and evaluating the RA in larger industrial settings would strengthen the external validity of the findings. Second, the current work focused on functional and structural quality attributes. It would contribute to a more complete picture of its trade-offs a dedicated investigation into the implications of the RA for non-functional attributes such as performance, security, and fault tolerance. Finally, the development of reference implementations for specific technology stacks would lower the adoption barrier for practitioners seeking to apply the proposed RA in their systems.

\appendix

\section{UML notation}\label{appendix:uml-ext}

\begin{figure}
  \centering
  \includegraphics[width=0.5\linewidth]{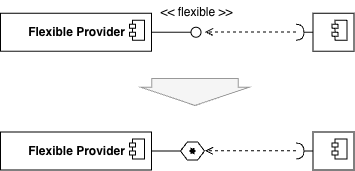}
  \caption{Flexible Provided Interface}
  \label{fig:fpi-notation}
\end{figure}

\begin{figure}
  \centering
  \includegraphics[width=0.5\linewidth]{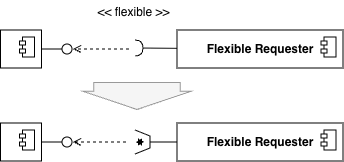}
  \caption{Flexible Requested Interface}
  \label{fig:fri-notation}
\end{figure}

\begin{itemize}
  \item \textbf{Flexible Provided Interface}: it extends the \underline{Provided Interface} notation by highlighting the flexibility in the incoming connection, i.e., it accepts multiple data structures from the clients; Fig. \ref{fig:fpi-notation} displays the notation design;
  
  \item \textbf{Flexible Requested Interface}: it extends the \underline{Requested Interface} by its flexibility when connecting out, i.e., it can handle multiple \underline{Provided Interfaces}; Fig. \ref{fig:fri-notation} displays the notation design.
\end{itemize}

\begin{figure}
  \centering
  \includegraphics[width=0.5\linewidth]{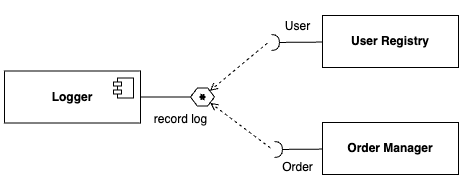}
  \caption{Example of Flexible Provided Interface}
  \label{fig:fpi-notation-example}
\end{figure}

\begin{figure}
  \centering
  \includegraphics[width=0.5\linewidth]{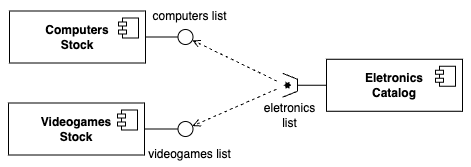}
  \caption{Example of Flexible Requested Interface}
  \label{fig:fri-notation-example}
\end{figure}

We provide two examples, one for each notation. In Fig. \ref{fig:fpi-notation-example}, a system containing the \textsc{User Registry}, \textsc{Order Management}, and \textsc{Logger} components, the Logger offers a \texttt{record log} operation as a \textbf{Flexible Provided Interface}. With that, the single operation can handle both the \texttt{User} and the \texttt{Order} records. In Fig. \ref{fig:fri-notation-example}, another system containing the \textsc{Computers Stock}, \textsc{Videogames Stock}, and \textsc{Eletronics Catalog} components, the aggregation of both electronics into a single list is exposed by a \textbf{Flexible Required Interface}: the \texttt{electronics list} operation. That is possible because of the flexibility in handling different interfaces.

\printcredits

\bibliographystyle{cas-model2-names}

\bibliography{cas-refs}



\end{document}